\documentclass[12pt,aps,preprint,pra]{revtex4}
\usepackage{graphicx}

\begin{document}

\title{A first principles investigated optical spectra of oxizided graphene}

\author{N. Singh$^{1, 2}$, T. P. Kaloni$^{1}$, and Udo Schwingenschl\"ogl$^{1}$}

\email{udo.schwingenschlogl@kaust.edu.sa}

\email{nirpendra.singh@kaust.edu.sa}

\affiliation{$^{1}$ Physical Science \& Engineering Division, KAUST, Thuwal 23955-6900, Kingdom of Saudi Arabia}

\affiliation{$^{2}$ Solar and Photovoltaic Energy Research Center (SPERC), KAUST, Thuwal 23955-6900, Kingdom of Saudi Arabia}

\begin{abstract}

The electronic and optical properties of mono, di, tri, and tetravacancies in graphene are studied in comparison to each other, using density functional theory. In addition, oxidized monovacancies are considered for different oxygen concentrations. Pristine graphene is found to be more absorptive than any defect configuration at low energy. We demonstrate characteristic differences in the optical spectra of the various defects for energies up to 3 eV. This makes it possible to quantify by optical spectroscopy the ratios of the defect species present in a sample.
\end{abstract}


\maketitle
While graphene is a zero band gap material, but a finite band gap is needed for various applications aiming at graphene based electronic
devices \cite{schwierz,kaloni3}. Functionalization is one of the possible methods to open a band gap for example by simple oxidation \cite{wu,priya,kaloni}. 
Graphene oxide (GO) with epoxy, carbonyl, or hydroxyl groups could allow to tune the a band gap and therefor tailor the electronic, mechanical,
and optical properties \cite{dai,andre}. The atomic structure of GO has been studied experimentally \cite{weiwei,ajayan} and theoretically \cite{sumit}. 
Recently, GO nanostructures have created a lot of attention due to the fact that it paved the way for solution based synthesis of graphene sheets, 
low cost, easy processibility, and compatibility with various substrates \cite{Joung}. The band gap of GO can be tunable by just varying the oxidation level. 
Fully oxidized GO can act as an electrical insulator and partially oxidized GO can act as a semiconductor \cite{Loh}. Moreover, experiments demonstrat 
that GO nanostructures have promising applications in photocatalysis \cite{Krishnamoorthy}. Reduction of GO may pave the way to mass production of graphene \cite{shenoy}.

While GO is usually insulating, a controlled deoxidation can lead to an electrically and optically active material that is transparent and
conducting. Furthermore, in contrast to pristine graphene, GO is fluorescent over a broad range of wavelengths, owing to its heterogeneous electronic
structure \cite{Loh}. It can contain different chemical compositions of carbon, oxygen, and hydrogen \cite{weiwei,lu}. While commonly hydroxyl and 
epoxy groups are found, there can be small contributions of carbonyl and carboxyl groups. Experimentally, a coverage of between 25\% and 75\% has 
been observed, reflecting that typically a quarter of the C$-$C bonds are double bonds whereas the rest are single bonds \cite{Katsnelson}. 
Adsorption behavior of oxygen atoms on the graphene sheets has been studied by using first-principles calculations \cite{ito} and found that the lattice 
constant increases with the increase of the ratio of $O/C$ because of the formation of the epoxy group. At 50\% $O/C$ ratio, a finite band gap of 3.39 eV is reported \cite{ito}.

The attachment of a carbonyl group leads to an almost planar $sp^{2}$ electronic configuration because the formation of C=O bonds induces little 
strain in the graphene sheet. On the contrary, the attachment of an epoxy group leads to a non-planar distorted $sp^{3}$ electronic configuration 
for those C atoms which are connected to O, which creates a significant strain on neighboring C$-$C bonds \cite{shenoy}. The combination of $sp^{2}$ 
and $sp^{3}$ configurations as well as defects breaks the hexagonal symmetry of pristine graphene and a band gap is opened. The coexistence of $sp^{2}$ and $sp^{3}$
configurations is confirmed experimentally \cite{yun} and theoretically \cite{shenoy}. The defects associated with dangling bonds enhance the reactivity substantially.

\begin{figure}
\includegraphics[width=1.00\textwidth,clip]{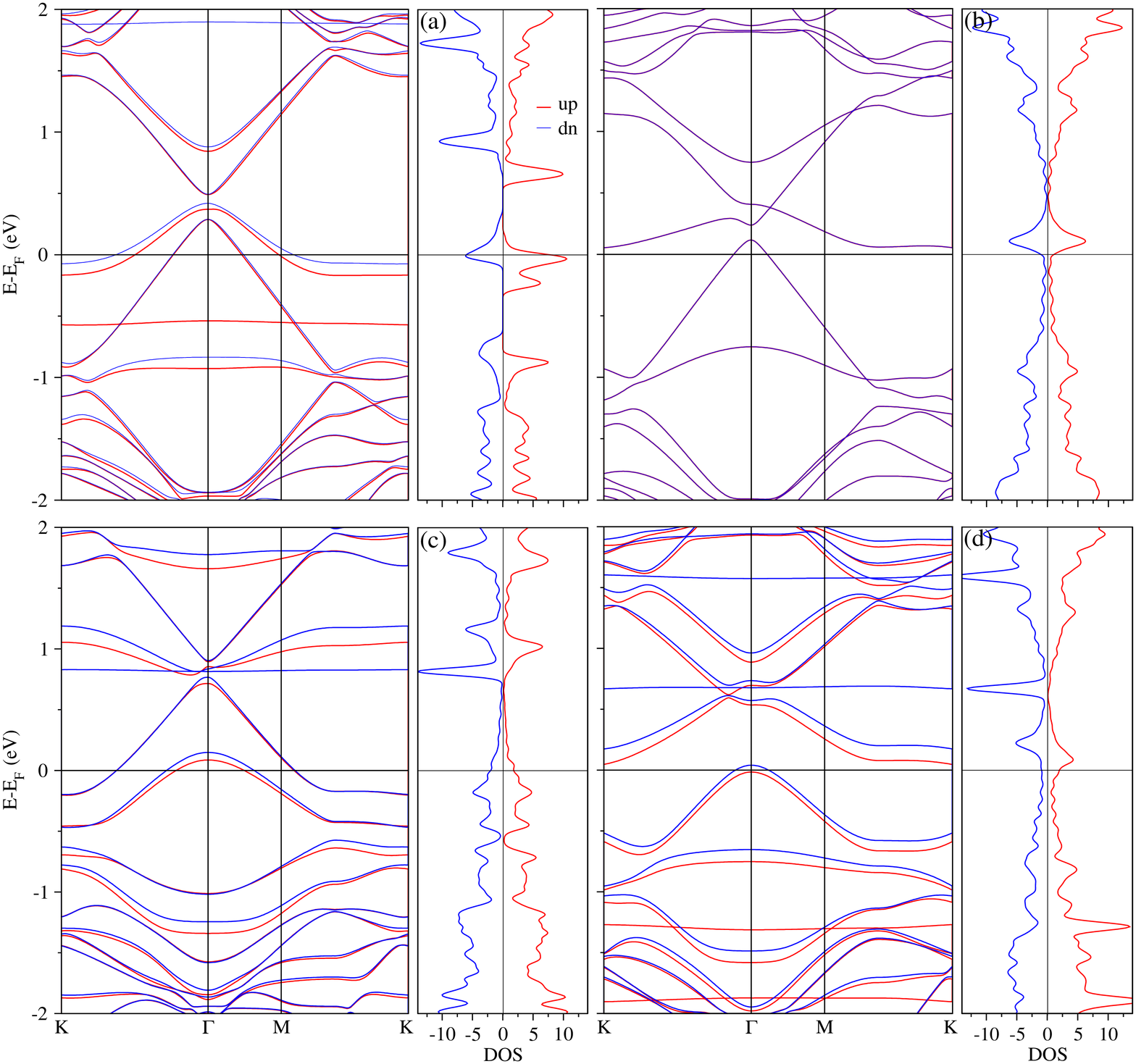}
\caption{Calculated band structures of (a) a mono-vacancy (b) a di-vacancy, (c) tri-vacancy, and (d) tetra-vacancy in graphene along the path 
K'$_1$ (0.6667 0.0000 0.0000), K$_2$ (0.3333 -0.5774  0.0000), $\Gamma$ (0.0000   0.0000  0.0000), K'$_1$ (0.6667 0.0000 0.0000), K$_1$ (0.3333  0.5773  0.0000), 
M$_1$ (0.500   0.2887  0.0000), and $\Gamma$ (0.0000   0.0000  0.0000) in unit of $\frac{2\pi}{a}$.}
\end{figure}

Recently, a theoretical investigation of the electronic and optical properties of GO (without vacancies) for different functional groups and various 
compositions has been reported in Ref \cite{priya}. The authors found that carbonyl groups are favourable for photoluminescense and that the optical 
gap of reduced GO is samaller than the optical gap of pristine and fully oxidize graphene. Theoretical and experimental studies \cite{Katsnelson1,Wang} 
indicate that hydroxyl, carboxyl, and other functional groups can easily be attached to vacancies in graphene than to pristine graphene. 
Therefore, a study of the electronic and optical properties of defective GO becomes critical. Optical properties of oxidized mono-, di-, tri-, and 
tetra-vacancies in graphene have not been reported so far. In this work, we use first principles calculations of to provide insight into this topic. 

Our calculations are based on density functional theory and carried out using the generalized gradient approximation of Quantum Espresso pacakage, \cite{paolo,pbe}. 
All calculations are performed with a plane wave cutoff energy of 544 eV. We use a Monkhorst-Pack \cite{Monk} of $8\times8\times1$ k-mesh for the Brillouin zone 
integration in order to relaxing the structures and achieving highly accurate electronic structure. We observe that a $5\times5$ supercell of graphene is sufficiently
large for monovacancies and oxidized monovacancies \cite{kaloni}. Our supercell has a lattice constant of $a=12.2$ \AA{}\ and extends $c=20$ \AA{} in the 
perpendicular direction. It has been reported that nearby vacancies behave independently when they are separated by $\sim$7 \AA{}\ \cite{andrey}.
Hence, we use a $6\times6$ supercell for di-, tri-, and tetra-vacancies to avoid artificial interaction of the periodic images. This supercell
has a lattice constant of $a=14.69$ \AA{}\ and again $c=20$ \AA{}. The cell parameters and atomic positions are fully relaxed until a force
convergence of 0.05 eV/\AA{} and an energy convergence of 10$^{-7}$ eV are reached. The relaxed structures are used to calculate the optical properties by 
WIEN2k \cite{Wien2k} code. For the optical calculations, a dense mesh of uniformely distributed \textbf{k}-points is required. Hence, the Brillouin zone integration 
is performed using tetrahedron method with 180 \textbf{k}-points in the irreducible part of the Brillouin zone. Well converged solutions are obtained for 
$R_{mt}\times K_{max}$ =7, where $K_{max}$ is the plane-wave cut-off and $R_{mt}$ is the smallest of all atomic sphere radii.

The dielectric function ($\varepsilon(\omega)=\varepsilon_1(\omega)+i\varepsilon_2(\omega)$) is known to describe the optical response of the medium. The interband contribution to the
imaginary part of the dielectric function $\varepsilon(\omega)$ is calculated by summing all transitions from occupied to unoccupied states (with fixed {\bf k}) over 
the Brillouin zone, weighted with the appropriate matrix elements giving the probability for the transitions. The imaginary part of dielectric function $\varepsilon_2(\omega)$ 
is given as in \cite{wooten} by
\begin{equation}
\epsilon_2(\omega) = \frac{4\pi^2e^2}{m^2\omega^2} \sum_{i,j}\int <i|M|j>^2\times f_i(1-f_i)\delta(E_f-E_i-\omega)d^3k
\end{equation}

Where M is dipole matrix, \emph{i} and $\emph{j}$ are the initial and final states, respectively, $\emph{f}_i$ is the Fermi distribution function for the $i_th$ 
states, and $E_i$ is the energy of electron in the $i_th$ state. The real part of the dielectric function can be extrated from the imaginary part using the 
Kramers-Kronig relation \cite{wooten,Yu} in the from 
\begin{equation}
\epsilon_1(\omega) = 1 + \frac{2}{\pi} P \int_0^{\infty} \frac{\omega^{'} \epsilon_2 (\omega^{'})}{\omega^{'2} - \omega^2} d\omega^{'}
\end{equation}

Where P implies the principle value of the integral. 

The reflectivity spectra are derived from Fresnel's formula for normal incidence assuming an orientation of the crystal surface parallel to 
the optical axis using the relation \cite{fox}

\begin{equation}
R(\omega)=|\frac{\sqrt{\varepsilon(\omega)}-1}{\sqrt{\varepsilon(\omega)}+1}|^2
\end{equation}

The knowledge of both real and imaginary parts of the dielectric tensor allows the calculations of the important optical functions. 
We calculate the absorption, the real part of optical conductivity, and the electron energy-loss spectrum using the following expressions \cite{fox,delin}

\begin{equation}
\alpha(\omega)=\sqrt{2}\omega(\sqrt{\varepsilon_1(\omega)^2+\varepsilon_2(\omega)^2}-\varepsilon_1(\omega))^{1/2}
\end{equation}

\begin{equation}
Re\sigma(\omega)=\frac{\omega\varepsilon_2}{4\pi}
\end{equation}

\begin{equation}
L(\omega)=\frac{\varepsilon_2(\omega)}{\varepsilon(\omega)^2+\varepsilon_2(\omega)^2}
\end{equation}

This approach has been successfully applied to narrow band gap materials including rare earth Zintl compounds \cite{Zintl, Zintl1}, 
and oxides \cite{singh}. For a reliable integration, a set of 180 \textbf{k}-points in the irreducible wedge of the Brillouin zone is used. A Lorentzian broadening 
is used to simulate the effects of finite life-time and finite resolution of the optical measurement.
\begin{figure}
\includegraphics[width=1.00\textwidth,clip]{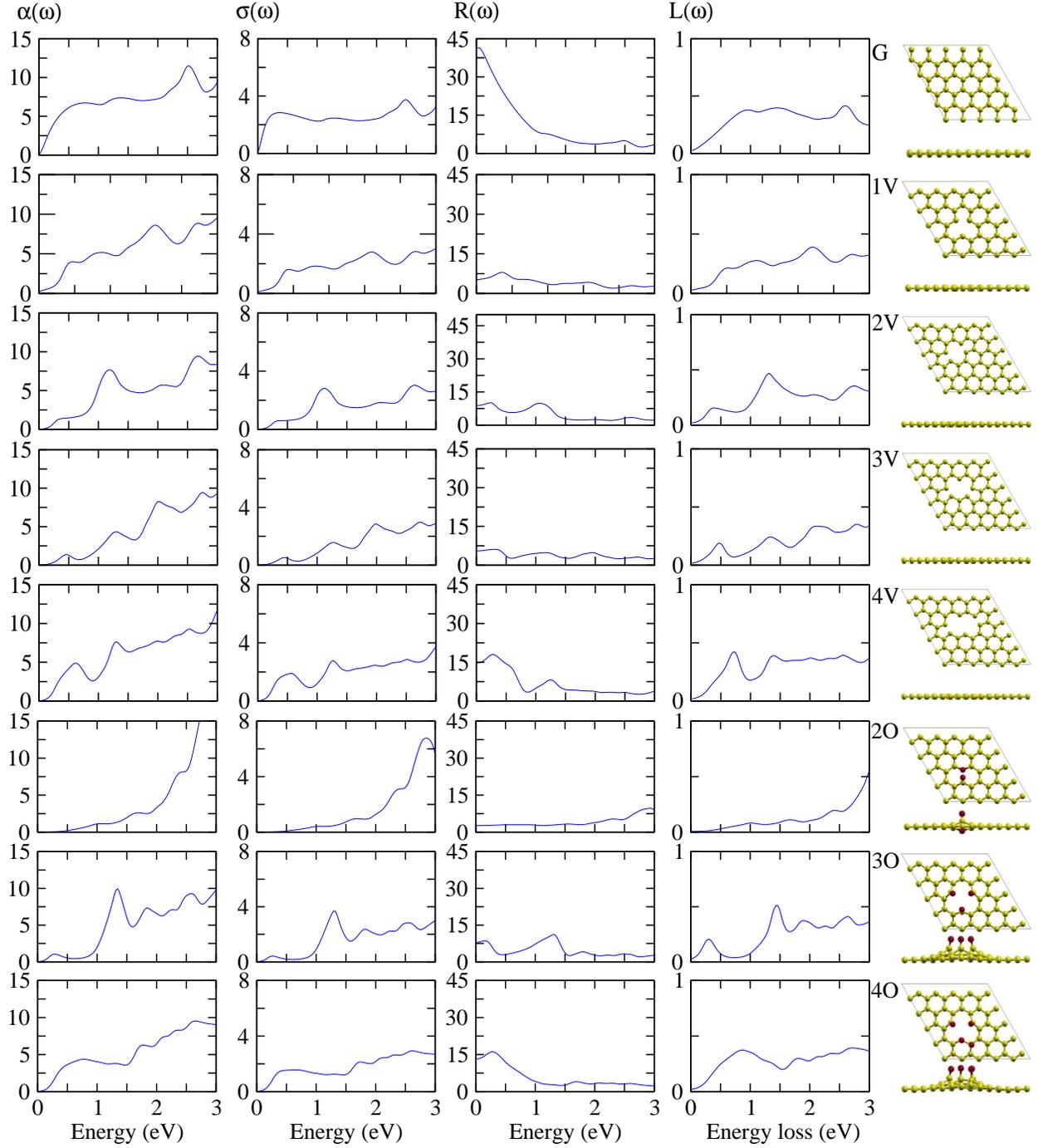}
\caption{Calculated optical absorption $\alpha(\omega)$ in $10^{4}$ cm$^{-1}$, optical conductivity $\sigma(\omega)$, reflectivity $R(\omega)$ in \%,
and energy loss function $L(\omega)$ of pristine graphene, as compared to graphene with mono, di-, tri, and tetra-vacancies and oxidized monovacancy.}
\end{figure}

The calculated values of formation energy of mono-, di-, tri-, and tetra-vacancies in graphene are 7.50, 6.94 eV, 11.45 eV, and 12.58 eV, respectively. 
This means the formation of a divacancy in graphene is more favourable than the formation of a single vacancy. Furthermore, a divacancy is known to be more stable than two 
isolated monovacancies (whose migration energy barrier is rather low), because the dangling C-C bonds of atoms next to the vacancy can be passivated by eachother \cite{Lee}. 
The vacancies in graphene induce ferromagnetism with total magnetic moments of 1.35 $\mu_B$, 1.00 $\mu_B$, 2.00 $\mu_B$ for mono-, tri-, tetra-vacancies, respectively. 
Di-vacancies shows no spin-polarization. The results of our band structure (BS) calculations for mono-, di-, tri-, and tetra-vacancies are shown in Fig.\ 1 
together with the corresponding DOS. For the sake of comparison we have included the BS (dotted lines) of pristine graphene. For an oxidized monovacancy, 
the magnetic and electronic properties have been reported in previous literature \cite{kaloni,dai,Yazyev}. In $6\times6$ supercell, the Dirac cone is 
shifted to the $\Gamma$-point due to Brillouin zone folding in $6\times6$ supercell. In case of the tetra-vacancy, the BS shows that a single minority spin 
band crosses the Fermi energy ($E_F$) at the $\Gamma$-point and leaves the system metallic, whereas for the di-, tri-vacancies both majority and minority bands 
cross $E_F$. Moreover, an upward shift of the Dirac point is indicative of a hole-doped system. In case of the di-vacancy, the DOS is identical for the majority 
and minority spins, reflecting spin-degeneracy. It means pristine graphene becomes ferromagnetic by a single vacancy defect and can be non-magnetic metal by divacancy.
 
The optical spectroscopy is a valuable tool in material science. Here, In the optical calculations, selfenergy 
and excitonic effects are not taken into account. It has been shown for graphene that in the energy range upto 3 eV, 
where the approximation of Dirac particles is valid, the influence of the many-particle effects is negligible. 
The calculated optical spectra of pristine graphene and its functionalized derivatives are addressed in Fig 1. 
The optical spectra show that for the two adsorbed O atoms, a band gap of 0.5 eV is opened due to the symmtery breaking and 
increased $sp^3$ characters. A semiconducting behaviour for this configuration is conformed by our previous calculations of BS and 
DOS \cite{kaloni}. As comparised to pristine graphene, all defects are found to create metallic states as shown in Fig 1.

In general, pristine graphene is more absorptive as compared to the other systems but for di-vacancy, absoption between 7.5 eV to 10 eV is higher than pristine graphene. 
The $\alpha(\omega)$ peak at 4.5 eV for pristine graphene splits into two peaks at 2.7 eV and 5 eV by the splitting of the Dirac cone for two attached O atoms. 
The splitting increasing with the O coverage. For di-vacancy and three adsorbed O atoms an additional sharp peak at 1.25 eV (visible region) is found which is absent for other systems. 
The pristine graphene have high reflectivity in the low energy as compared to other cases. The reflectivity is higher for the tetra-vacancy (most pronounced metallicity) 
as compare to mono-, di-, and tri-vacancies, but lower than for pristine graphene. The reflectivity in low energy range is the lowest for the case of monovacancy with 
two attached O atoms, due to its semiconducting nature and increases again for three and four attached O atoms. The most prominent peaks in $\sigma(\omega)$ become broaden and 
the magnitude also decreases as one moves from pristine graphene to tetra-vacancies. The $\sigma(\omega)$ is low for tri-, tetra-vacancies and monovacancy with four attached O atoms 
as compared to remaining systems. A large peak is observed at around 4.6 eV in all optical spectra for all the systems which is attributed to the $\pi-\pi^*$ transitions of the aromatic C$-$C atoms. 
The maximum peak in energy loss spectra is at 4.9 eV, which is assigned to the energy of the volume plasmon $\hbar$ $\omega_{p}$. This maximum peak positions remain same for all systems. 
The peaks in optical spectra originates from the transition from valence band to conduction band. The dominating peak in the energy loss spectrum broadens from graphene to the 
oxidize vacancies with monovacancy.  

In conclusion, we have studied the optical properties of graphene derivatives (clean and oxidized vacancies) by means of density functional theory. We find that the formation of divacancies in graphene is energetically favourable. Divacancies are also exceptional in the sense that they do not lead to a local magnetic moment. Mono, di, tri, and tetravacancies are found to be metallic, while an oxidized monovacancy with two adsorbed oxygen atoms leads to a band gap of 0.5 eV (due to a splitting of the Dirac cone). Our optical spectra show that pristine graphene has the highest absorption in the energy range below 2.5 eV. In two cases (tetravacancy and monovacancy with three adsorbed oxygen atoms) a prominent absorption peak appears in the visible range. Our calculations suggest that the types of (oxidized) defects present in a graphene sample can be quantified by optical spectroscopy. With this knowledge, the electronic and optical properties of graphene derivatives can be tuned by controlled oxidation and reduction.

\end{document}